\documentclass{ws-ijmpe}

\begin{document}

\markboth{Thomas J. Humanic}{Signatures for Black Hole production}

%
\catchline{}{}{}{}{}
%

\title{Signatures for Black Hole production from hadronic observables at the Large Hadron Collider}

\author{Thomas J. Humanic}

\address{Department of Physics, The Ohio State University\\
Columbus, Ohio, 43210
\\humanic@mps.ohio-state.edu}

\author{Benjamin Koch and Horst St{\"{o}}cker}

\address{Institut f{\"{u}}r Theoretische Physik, J. W. Goethe-Universit{\"{a}}t\\
Max von Laue Strasse 1, 60438 Frankfurt am Main, Germany \\
koch@th.physik.uni-frankfurt.de/stoecker@th.physik.uni-frankfurt.de}

\maketitle

\begin{abstract}
The concept of Large Extra Dimensions (LED) provides a way of
solving the Hierarchy Problem which concerns the weakness of gravity
compared with the strong and electro-weak forces. A consequence of
LED is that miniature Black Holes (mini-BHs) may be produced at the
Large Hadron Collider in $p+p$ collisions. The present work uses the
CHARYBDIS mini-BH generator code to simulate the hadronic signal
which might be expected in a mid-rapidity particle tracking detector
from the decay of these exotic objects if indeed they are produced.
An estimate is also given for Pb+Pb collisions.
\end{abstract}

\section{Introduction}
The Hierarchy Problem in particle physics is concerned with the
question of why gravity is so weak compared with the other forces in
nature, or posed another way, why the scale for gravity, i.e. the
Planck mass at ${\sim}10^{19}$ GeV, is so much larger than the
scales for the other forces in nature, the strong force with a Lund
string model scale of ${\sim}1$ GeV/fm and the electro-weak force
characterized by the mass of the W and Z bosons, ${\sim}100$ GeV.
Several solutions have been proposed to solve this problem, such as
1) Supersymmetry in which bosons and fermions are symmetric and
which unifies the strong and electro-weak forces at a scale just
below the Planck scale, 2) String Theory in which elementary
particles are represented by higher dimensional strings, e.g. 11
dimensional, and which is a theory of quantum gravity thought to be
valid up to the Planck scale and beyond, and 3) the concept of Large
Extra Dimensions (LED) which also assumes space-time has a higher
dimensionality than the normal $3+1$
dimensions\cite{add1998,rs1999}. The present study will be carried
out in the framework of the LED model of Arkani-Hamed, Dimopoulos,
and Dvali\cite{add1998,add1999}, called the ADD model, so some
further discussion of the model is given. This model has produced a
great deal of interest in the literature, and there exist many
papers that discuss it and its consequences, several of which are
referenced here\cite{gt2002,banks1999,landsberg2001,rizzo2004}.

The four main assumptions of the ADD model are: 1) space-time is
higher dimensional, so introduce $n$ extra spacial dimensions beyond
our usual three such that space-time is $3+1+n$ dimensional, 2) only
allow gravity, i.e. gravitons, to propagate in all $3+1+n$
dimensions, 3) assume the extra dimensions are ``compact'', i.e.
finite, so they are too small to be normally detected but large
enough to impact physics, and 4) assume that Standard Model
particles, e.g. quarks, gluons, photons, etc..., are confined to a
$3+1$ dimensional ``wall'' or ``brane'' representing our normal
world embedded in the higher dimensional space. The mechanism used
in the ADD model to solve the Hierarchy Problem is to define that it
doesn't exist: the reason gravity looks so weak in our $3+1$
dimensional world is that its force is diluted by existing in
$3+1+n$ dimensions, so the higher dimensional ``true'' Planck mass,
$M_P$, is much lower than the ``apparent'' Planck mass we measure in
our world. As will be shown below, by adjusting the number of extra
dimensions and their size, the higher dimensional Planck mass can be
brought down to a level low enough, i.e. ${\sim}1$ TeV, to eliminate
the Hierarchy Problem.

In addition to resolving the Hierarchy Problem, the ADD model leads
to consequences which may be observed in nature. One of these is due
to the compactification of the extra dimensions resulting in
``towers'' of new ``particle-in-a-box'' energy states called
Kaluza-Klein states (named after the researchers who in the 1930's
made an unsuccessful attempt to unify gravity with the
electromagnetic force using extra dimensions)\cite{rizzo2004}.
Kaluza-Klein states can be associated with a spectrum of graviton
states which could influence hard scattering processes at the Large
Hadron Collider (LHC) and could be Dark Matter
candidates\cite{add1999}. Another exciting consequence of the ADD
model is that miniature Black Holes could exist at the greatly
reduced Planck mass of around 1 TeV and thus might be produced in
TeV-scale particle collisions, such as at the
LHC\cite{gt2002,banks1999,landsberg2001}. Of course, in order for
the ADD model to be a viable solution to the Hierarchy Problem, none
of its consequences are allowed to conflict with existing
observations. It can be shown that for $n>2$ and $M_P{\sim}1$ TeV,
the ADD model does not conflict with existing astrophysical
observations, cosmology, or particle accelerator
experiments\cite{add1999} (some of these arguments will be presented
later in this note). A very interesting feature of the ADD model is
that it can be shown to be consistent with Type I String
Theory\cite{add1998}, which is characterized by extra dimensions,
open ended strings (SM particles) stuck to a $3+1$ dimensional
brane, and closed strings (gravitons) which can move freely in the
extra dimensions. Because of this, detection of unique signatures
predicted by the ADD model would be a strong suggestion of the
validity of String Theory and thus we might be able to
experimentally study String Theory effects at the 1 TeV
scale\cite{witten2002}!

In this spirit, the main goal of the present work is to study the
possibility of detecting unique signatures of miniature Black Holes
which may be created in LHC $p+p$ collisions. An estimate for
$Pb+Pb$ collisions is also made at the end. More specifically,
quantitative calculations will be carried out for mini-BH production
at the LHC using a code based on the ADD model,
CHARYBDIS\cite{harris2003}, and comparisons of charged hadron
production from mini-BH decay will be made with charged hadron
production from the PYTHIA QCD code\cite{pythia} in a particle
tracking detector whose acceptance is around mid-rapidity such as in
the ALICE experiment\cite{alice}. Since the concepts of Large Extra
Dimensions and Black Holes are less familiar than other aspects of
LHC physics, the interested reader can find some qualitative
discussions of these concepts elsewhere\cite{humanic2005}.

\section{Quantitative calculations of mini-BH production at the
LHC and detection via the hadronic channel} This section describes
quantitative calculations for BH production and detection using a
hadronic detector system with the performance and acceptance
characteristics of in the LHC ALICE experiment. Unless otherwise
noted, results presented refer to LHC $p+p$ collisions. Similar
studies have been carried out by others for the LHC ATLAS
experiment\cite{atlas2003,harris2004}. In the present study the BH
event generator code CHARYBDIS\cite{harris2003} is used. CHARYBDIS
represents a quantitative treatment of BH formation and decay for
$p+p$ collisions. Three of its main features are 1) it integrates
over parton distribution functions, 2) it calculates BH decay
incorporating ``grey body'' factors, and 3) it is coupled to the
PYTHIA QCD code\cite{pythia} for parton evolution and hadronization.
Parton distribution functions used in this study are the same as
used by Dimopoulos and Landsberg\cite{landsberg2001}.

Figure 1 shows the production cross section and number of BHs
produced in a one-year running period at the LHC for $p+p$ at a
luminosity of $10^{34}$ cm$^{-2}$s$^{-1}$, and similar numbers for
Run I at the Tevatron, as a function of the Planck mass from
CHARYBDIS+PYTHIA. As seen, at full luminosity at the LHC for one
year there are about $10^9$ BHs produced if the Planck mass is 1
TeV, exponentially decreasing to only 1 BH being produced in a year
for 10 TeV. For the Tevatron, it is seen that only a few BHs are
expected to have been produced during Run I if the Planck mass were
exactly at 1 TeV, dropping to less than one BH for 1.1 TeV. Thus, it
is not suprising that no evidence of BH formation has been reported
so far at the Tevatron.

\begin{figure}
\begin{center}
\hspace{1.3in} \scalebox{.6}{\includegraphics{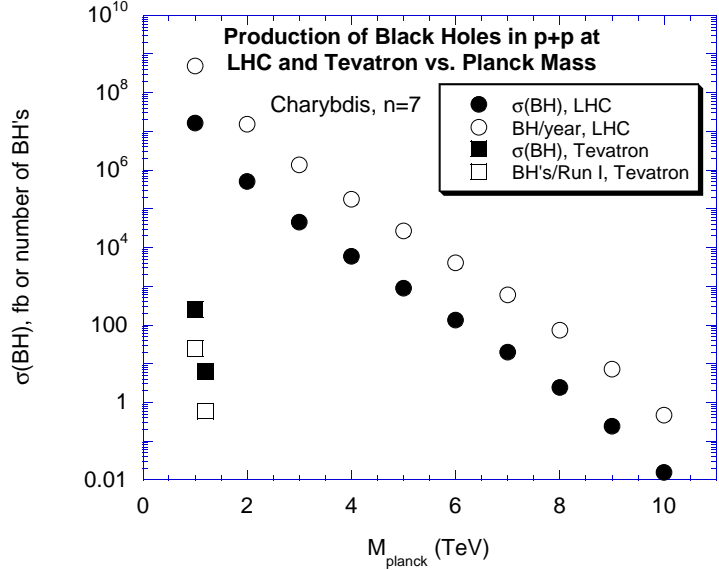}}
\caption{Cross section and yield for mini-BH production vs. $M_P$
for the LHC (full luminosity and energy) and Tevatron (Run I) using
CHARYBDIS.} \label{fig:bh10}
\end{center}
\end{figure}

Figure 2 shows a comparison between transverse momentum
distributions for charged hadrons at mid-rapidity at full LHC energy
from CHARYBDIS+PYTHIA for two values of $M_P$ with that from PYTHIA
alone, labeled ``QCD only'' (this plot is patched together from
various PYTHIA runs with increasing values for the hardness of the
$2\rightarrow 2$ parton collision, each run represented by a
different color). As seen for $M_P=1$ TeV, hadrons from BH decays
dominate over QCD processes for $p_T>100$ GeV/c, whereas for $M_P=5$
TeV, BH decays only become important for $p_T>1.2$ TeV. Since the
ALICE experiment does not presently forsee having large-acceptance
calomimetry capable of accurate particle $p_T$ measurements to very
high values, alternative observables which are sensitive to the
possibility of $M_P>1$ TeV are needed. Taking advantage of the
large-acceptance precision tracking detectors available in ALICE,
namely the combined Inner Tracking System (ITS), Time Projection
Chamber (TPC) and Transition Radiation Detector (TRD) tracking, two
event-by-event observables look promising for BH studies: charged
multiplicity and summed $p_T$. The particle acceptance for charged
multiplicity in rapidity and $p_T$ is represented by $-2<y<2$ and
$p_T>0.1$ GeV/c, respectively. A reasonable acceptance which can be
taken for summed $p_T$ per event is represented by $-0.9<y<0.9$ and
$0.1<p_T<300$ GeV/c. These observables will be studied below.

\begin{figure}
\begin{center}
\hspace{1.3in} \scalebox{.6}{\includegraphics{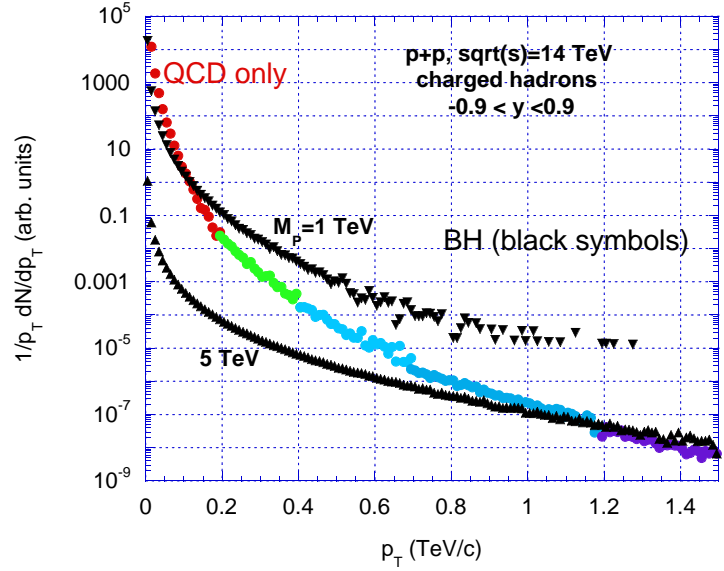}}
\caption{Transverse momentum distributions for charged hadrons from
BH decay (CHARYBDIS) compared with $2\rightarrow2$ hard QCD
processes (PYTHIA). } \label{fig:bh11}
\end{center}
\end{figure}

The strategy of the present study will be the following: a) compare
charged hadron production from BHs using CHARYBDIS+PYTHIA with QCD
high-$p_T$ processes from PYTHIA alone, b) detect these hadrons in
the ALICE ITS+TPC+TRD tracking acceptance with momentum resolution
effects included and c) try to find a simple triggering scheme to
use for this. For all of the plots shown below for the ALICE study,
the LHC will be assumed to give $\sqrt{s}=14$ TeV $p+p$ collisions
at a ``year-1'' luminosity of $10^{31}$ cm$^{-2}$s$^{-1}$. A value
for the number of extra dimensions will be taken to be $n=7$. Note
that for a given $M_P$ the results from CHARYBDIS do not depend
sensitively on the exact value of $n$ used for values of this order,
e.g. for $n=6$ the BH production cross section is within 10\% of the
$n=7$ case. The charged hadrons included in all plots are pions,
kaons, protons, and their anti-particles. The ALICE tracking
acceptance is simulated with rapidity and transverse momentum cuts,
and momentum resolution effects are conservatively simulated
assuming $\Delta p/p=0.16 p$, where $p$ is the particle momentum.

Figures 3 and 4 show plots of charged hadron multiplicity per event
and summed $p_T$ per event, respectively, for BH and QCD events in
the ALICE acceptances for running a minimum-bias trigger for four
months of initial LHC luminosity. The maximum data acquisition rate
for $p+p$ in ALICE is taken to be 100 Hz. For these running
conditions, it is seen that only a few BH events will be visible
above the QCD background and only for $M_P=1$ TeV, occuring for
multiplicities greater than 200 and summed $p_T$ greater than 0.5
TeV/c.

\begin{figure}
\begin{center}
\hspace{1.3in} \scalebox{.6}{\includegraphics{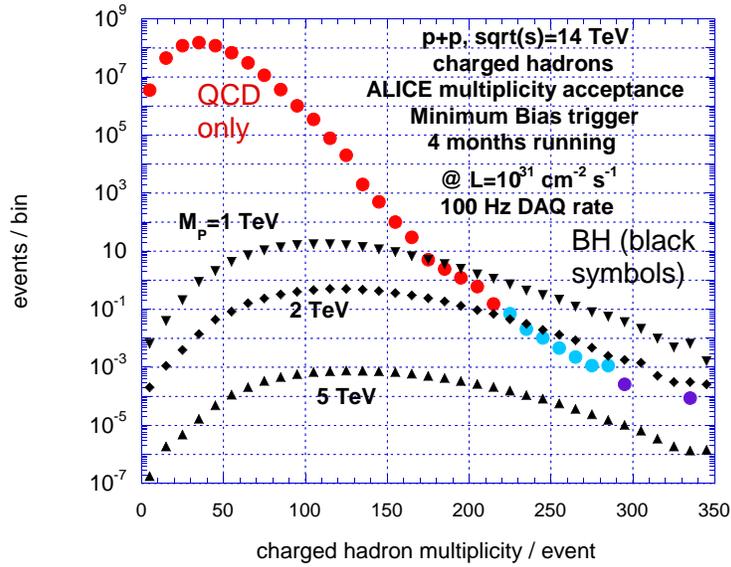}}
\caption{Multiplicity distributions for charged hadrons for minimum
bias triggering in ALICE.} \label{fig:bh12}
\end{center}
\end{figure}

\begin{figure}
\begin{center}
\hspace{1.3in} \scalebox{.6}{\includegraphics{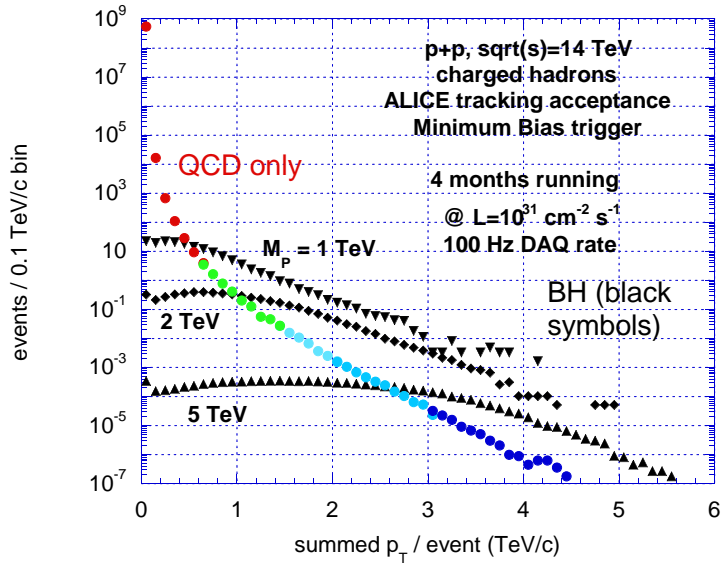}}
\caption{Summed $p_T$ per event distributions for charged hadrons
for minimum bias triggering in ALICE} \label{fig:bh13}
\end{center}
\end{figure}

In order to improve this situation, assume it is possible to apply a
simple charged multiplicity trigger to ALICE events in the tracking
acceptance $-0.9<y<0.9$ where only events with multiplicity greater
than 65 are accepted. Possible detectors which could be used to
implement such a trigger are the TRD and ITS, but further study of
the details of this is required. Figures 5 and 6 show plots
corresponding to Figures 3 and 4 above except with this multiplicity
trigger. The data rate for this trigger for the LHC luminosity used
is only 1 Hz, which is 1\% of the maximum data acquisition rate for
ALICE $p+p$ data and which would thus have only a small impact on
the overall data-taking rate for ALICE. As expected, this trigger is
seen to greatly reduce the QCD background allowing for significant
BH signals to be detected during this short running period. For
charged multiplicity, the sensitivity to $M_P$ is raised to 2 TeV
and hundreds of BH events above background corresponding to this
case are seen for multiplicities greater than 250. The situation is
seen to be even better for summed $p_T$, where tens of thousands of
BH events are seen above background for the $M_P=1$ TeV case, and
tens of BH events may be seen above background even for the $M_P=5$
TeV case in this running period. The signature for BH creation from
these simple observables is seen to be an abrupt flattening of the
slope of either distribution as the transition from pure QCD to BH
dominated charged particle production takes place. For $M_P<2$ TeV,
this flattening should be seen in both distributions in ALICE giving
a redundancy to this signature. The point in multiplicity and/or
summed $p_T$ where the flattening occurs would be related to the
value of $M_P$ which in principle could then be determined.

\begin{figure}
\begin{center}
\hspace{1.3in} \scalebox{.6}{\includegraphics{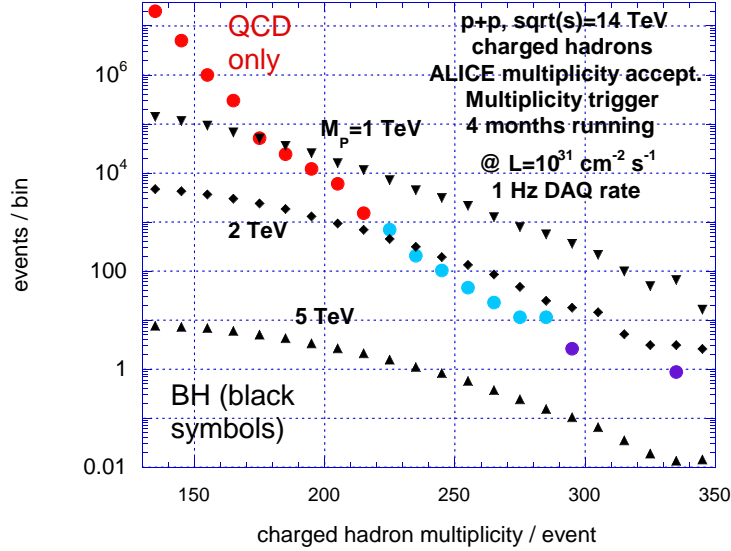}}
\caption{Multiplicity distributions for charged hadrons using a
multiplicity trigger in ALICE.} \label{fig:bh14}
\end{center}
\end{figure}

\begin{figure}
\begin{center}
\hspace{1.3in} \scalebox{.6}{\includegraphics{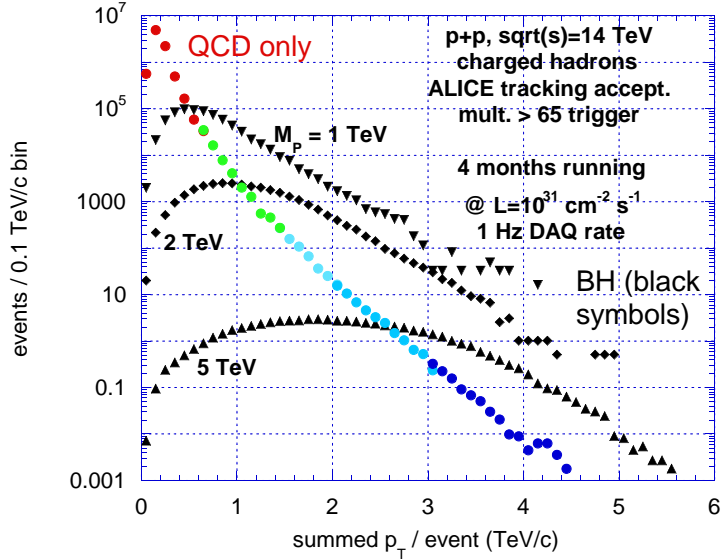}}
\caption{Summed $p_T$ per event distributions for charged hadrons
using a multiplicity trigger in ALICE.} \label{fig:bh15}
\end{center}
\end{figure}

It is also interesting to look at other hadronic observables which
might be used as signatures for BH production and could be
correlated with signals seen in the multiplicity and summed $p_T$
observables discussed above. One of these is the
disappearance/suppression of high $p_T$ di-jets if BHs are
produced\cite{stoecker2006}. Since BHs are thought to decay more or
less isotropically by sequential Standard Model particle emissions,
one would expect to observe an enhancement of mono-jets over the
usual back-to-back di-jet production seen in $p+p$ collisions. One
way to study this is to look at $dN/d(\Delta \phi)$ hadronic
distributions, where $\Delta \phi$ is the difference in azymuthal
coordinate between two hadrons in a given event. For the usual
back-to-back di-jet production one expects peaks in this
distribution at $\Delta \phi$ of 0 and $\pi$, whereas for mono-jets
one would only see a peak at 0. Figure 7 shows a preliminary study
of $dN/d(\Delta \phi)$ in the ALICE acceptance comparing BH events
(CHARYBDIS) and QCD events (PYTHIA) for several different
kinematical regions. A suppression of the backward peak is observed
for the BH events and is seen to depend on the kinematical region
used.

\begin{figure}
\begin{center}
\hspace{1.3in} \scalebox{.8}{\includegraphics{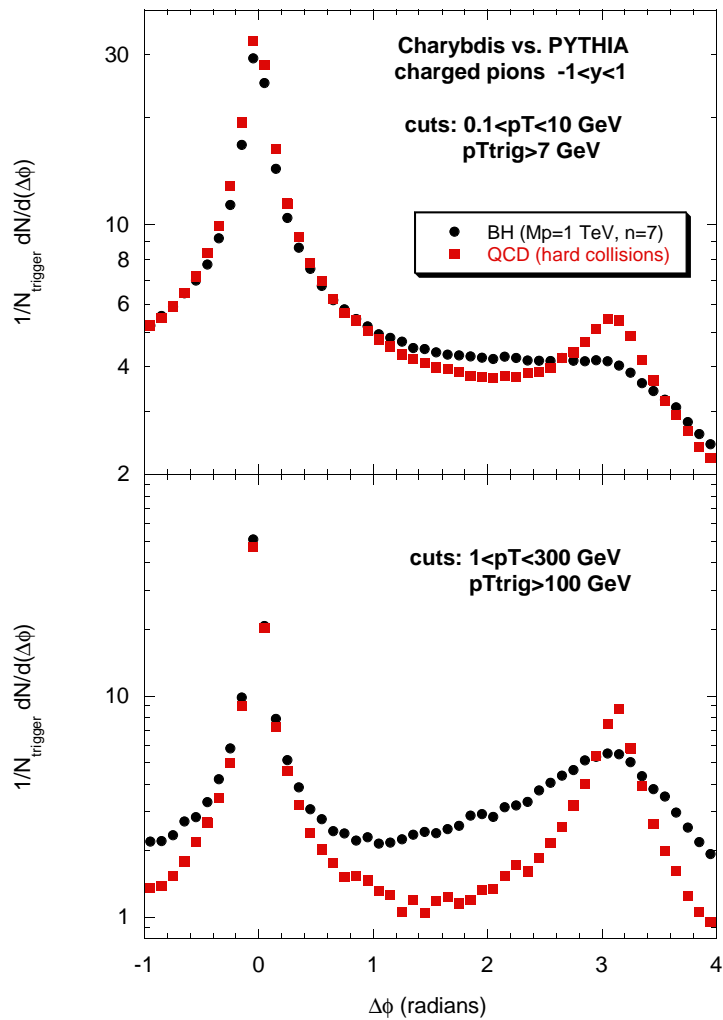}}
\caption{$\Delta \phi$ plots for the ALICE tracking acceptance for
simulated BH and QCD events for two kinematic regions.}
\label{fig:dndphi}
\end{center}
\end{figure}

Another possible signature is the direct detection of BH relics
which are BHs left over at the end of the Hawking decay of mass near
the higher-dimensional Planck Mass ($\sim 1$ TeV) which may be
stable\cite{sabine2004,sabine2005,bonanno2006}. A preliminary study
of the ability of ALICE to detect charged BH relics of 2 TeV mass is
shown in Figure 8. The charged BH relic formation was simulated
using a modified version of the CHARBYDIS code\cite{Koch2005}.
Acceptance, momentum resolution and TOF resolution effects are
included in this study. As seen, a 2 TeV charged relic is easily
observed in the ALICE detector.

\begin{figure}
\begin{center}
\hspace{1.3in} \scalebox{.6}{\includegraphics{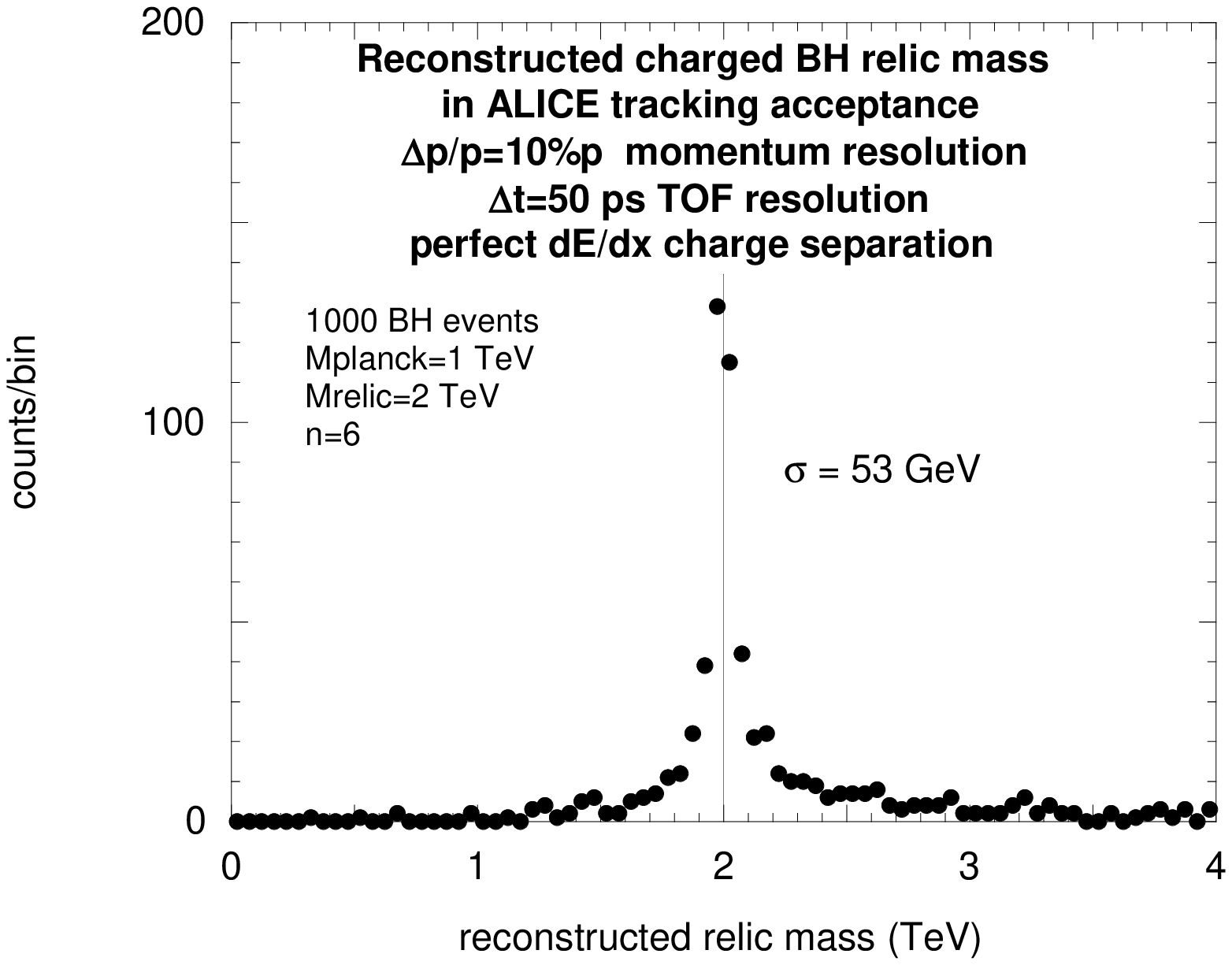}}
\caption{Mass plot for charged relic reconstruction in ALICE.
Charged BH relics with a mass of 2 TeV were simulated for this
study.} \label{fig:mrelic}
\end{center}
\end{figure}

The final topic which will be discussed is the ability to
produce/detect BHs in LHC $Pb+Pb$ running, which will take place
with lower energy, luminosity, and runtime per year compared with
$p+p$. As a lower-limit estimate, assume that a $Pb+Pb$ collision
can be represented as a superposition of free $p+p$ collisions with
high $p_T$ $2\rightarrow 2$ QCD processes. Under this assumption
PYTHIA and CHARYBDIS can be used. Calculate the effective $p+p$
luminosity, $L_{eff}$, from the expected $Pb+Pb$ luminosity of
$L_{PbPb} = 10^{27}$ cm$^{-2}$s$^{-1}$, assuming 5\% centrality
collisions. Coincidentally, it is found that $L_{eff}\sim 10^{31}$
cm$^{-2}$s$^{-1}$, the same luminosity taken in the above for ALICE
p+p running! Figure 9 shows the summed $p_T$ distribution for
charged hadrons simulated in this way for $Pb+Pb$ collisions as
detected in the ALICE acceptance. Only hadrons produced with $p_T
>10$ GeV/c are included in order to minimize the background from
soft processes which dominate the heavy-ion collision. As seen, for
four months of $Pb+Pb$ running (which represents four {\it years} of
LHC running since $Pb+Pb$ will only be run one month per year) the
case for $M_P=1$ TeV can easily be detected above the QCD
background, whereas for $M_P=2$ TeV detection is already becoming
problematic.

\begin{figure}
\begin{center}
\hspace{1.3in} \scalebox{.6}{\includegraphics{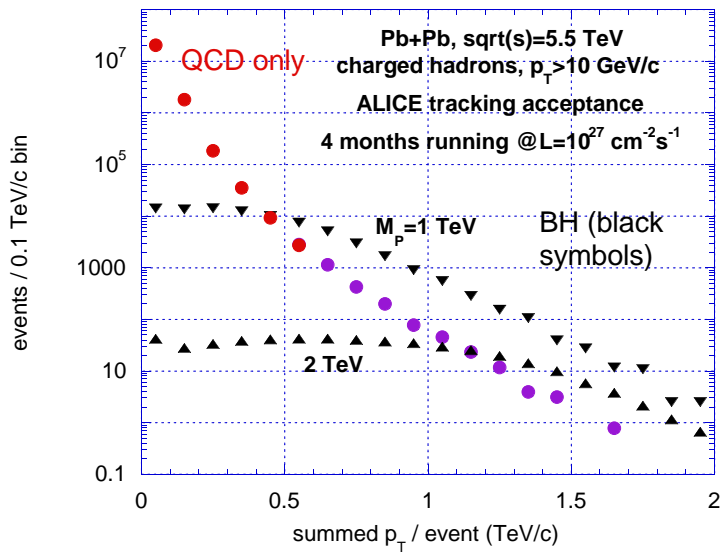}}
\caption{Summed $p_T$ per event distributions for charged hadrons
simulated for $Pb+Pb$ collisions with a central trigger in ALICE.}
\label{fig:pbept}
\end{center}
\end{figure}

\section{Summary}
Models with Large Extra Dimensions have exciting consequences
including the possible creation of mini-BHs at the LHC. Under the
proper conditions experiments such as the ALICE experiment have the
capability to detect BHs from charged hadronic observables for
higher-dimensional Planck masses ranging from 1 to 5 TeV within the
first four months of LHC running. A simple method for triggering on
charged particle multiplicity to enhance the BH signal is suggested.

\section{Acknowledgements}
The authors are grateful to Chris Harris for his help in running his
code CHARYBDIS. One of the authors (Humanic) would also like to
acknowledge the National Science Foundation for supporting his LHC
work under grant number PHY-0355007.


\begin{thebibliography}{99}

\bibitem{add1998}N. Arkani-Hamad,S. Dimopoulos,and G. Dvali,
{\it Phys.Lett.} {\bf B429}, 263 (1998).

\bibitem{rs1999}L. Randall and R. Sundrum, {\it Phys.Rev.Lett.}
{\bf 83}, 3370 (1999).

\bibitem{add1999}N. Arkani-Hamad,S. Dimopoulos,and G. Dvali,
{\it Phys.Rev.D} {\bf 59}, 086004 (1999).

\bibitem{gt2002}Steven B. Giddings and Scott Thomas,
{\it Phys.Rev.D} {\bf 65}, 056010 (2002).

\bibitem{banks1999}T.~Banks and W.~Fischler, arXiv:hep-th/9906038
(1999).

\bibitem{landsberg2001}S.~Dimopoulos and G.~Landsberg, {\it Phys.Rev.Lett.} {\bf 87},
161602 (2001).

\bibitem{rizzo2004}T. G. Rizzo, arXiv:hep-ph/0409309v1 (2004).

\bibitem{witten2002}E. Witten, arXiv:hep-th/0212247v1 (2002).

\bibitem{harris2003}C. M. Harris, P. Richardson, and B. R. Webber,
arXiv:hep-ph/0409309v1 (2004).

\bibitem{pythia}T. Sjostrand, L. Lonnblad and S Mrenna, PYTHIA 6.2
Physics and Manual, arXiv:hep-ph/0108264 (2001).

\bibitem{alice}ALICE Collaboration, F. Carminiti et al, {\it J.
Phys.G:Nucl.Part.Phys.} {\bf 30}, 1517 (2004).

\bibitem{humanic2005}Thomas J. Humanic, ALICE internal note:
ALICE-INT-2005-017.

\bibitem{atlas2003}J. Tanaka, T. Yamamura, S. Asai, and J. Kanzaki,
ATLAS internal note: ATL-PHYS-2003-037 (2003).

\bibitem{harris2004}C. M. Harris, M. J. Palmer, M. A. Parker,
P. Richardson, A. Sabertfakhri and B. R. Webber,
arXiv:hep-ph/0411022v1 (2004).

\bibitem{stoecker2006}H. St{\"{o}}cker, arXiv:hep-ph/0605062 (2006);
S.~Hofmann, M.~Bleicher, L.~Gerland, S.~Hossenfelder, S.~Schwabe and
H.~St{\"{o}}cker, arXiv:hep-ph/0111052 (2001); S.~Hossenfelder,
M.~Bleicher and H.~Stoecker, Int.\ J.\ Mod.\ Phys.\ D {\bf 13}
(2004) 1453.

\bibitem{sabine2004}S. Hossenfelder, M. Bleicher and H. St{\"{o}}cker,
arXiv:hep-ph/0405031 (2004).

\bibitem{sabine2005}S.~Hossenfelder, S.~Hofmann, M.~Bleicher and
H.~St{\"{o}}cker, {\it Phys.Rev.D}{\bf 66}, 101502 (2002); S.
~Hossenfelder, S.~Hofmann, M.~Bleicher and H.~St{\"{o}}cker, {\it
Phys.Lett.} {\bf B566}, 233 (2003).

\bibitem{bonanno2006}A. Bonanno and M. Reuter, {\it Phys.Rev.D} {\bf
73}, 083005 (2006).

\bibitem{Koch2005}
B.~Koch, M.~Bleicher and S.~Hossenfelder,
JHEP {\bf 0510} (2005) 053 [arXiv:hep-ph/0507138]; S.~Hossenfelder,
B.~Koch and M.~Bleicher,
arXiv:hep-ph/0507140.

\end{thebibliography}
\end{document}